\documentclass[aps,amsmath,amssymb,twocolumn,showpacs]{revtex4-1}
\usepackage{graphicx}
\usepackage{dcolumn}
\usepackage{bm}
\usepackage{color}

\begin{document}

\title{Thermal conductance and thermoelectric figure of merit \\ of
  C$_{60}$-based single-molecule junctions: electrons, phonons, and photons}

\author{J. C. Kl\"ockner$^{1}$}
\email{Corresponding author: Jan.Kloeckner@uni-konstanz.de}
\author{R. Siebler$^{1}$}
\author{J. C. Cuevas$^{2}$}
\author{F. Pauly$^{1}$}

\affiliation{$^{1}$Department of Physics, University of Konstanz, D-78457
  Konstanz, Germany}

\affiliation{$^{2}$Departamento de F\'{\i}sica Te\'orica de la Materia
  Condensada and Condensed Matter Physics Center (IFIMAC), Universidad
  Aut\'onoma de Madrid, E-28049 Madrid, Spain}

\date{\today}

\begin{abstract} 
Motivated by recent experiments, we present here an \emph{ab initio} study of
the impact of the phonon transport on the thermal conductance and
thermoelectric figure of merit of C$_{60}$-based single-molecule junctions.
To be precise, we combine density functional theory with nonequilibrium
Green's function techniques to compute these two quantities in junctions with
either a C$_{60}$ monomer or a C$_{60}$ dimer connected to gold electrodes,
taking into account the contributions of both electrons and phonons. Our
results show that for C$_{60}$ monomer junctions phonon transport plays a
minor role in the thermal conductance and, in turn, in the figure of merit,
which can reach values on the order of 0.1, depending on the contact geometry. 
In C$_{60}$ dimer junctions, phonons are transported less efficiently, but they 
completely dominate the thermal conductance and reduce the figure of merit as 
compared to monomer junctions. Thus, claims that by stacking C$_{60}$ molecules 
one could achieve high thermoelectric 
performance, which have been made without considering the phonon contribution, 
are not justified. Moreover, we analyze the relevance of near-field thermal 
radiation for the figure of merit of these junctions within the framework of 
fluctuational electrodynamics. We conclude that photon tunneling can be another 
detrimental factor for the thermoelectric performance, which has been overlooked 
so far in the field of molecular electronics. Our study illustrates the crucial 
roles that phonon transport and photon tunneling can play, when critically assessing
the performance of molecular junctions as potential nanoscale thermoelectric
devices.
\end{abstract}

\pacs{}


\maketitle

\section{Introduction}

Molecular junctions have turned out to be ideal systems to explore and
establish the fundamental principles that govern charge and energy transport
at the nanoscale \cite{Cuevas2010,Cui2017b}. In particular, recent
experimental advances have made possible to study key aspects of energy and
heat conduction in molecular junctions such as thermoelectricity
\cite{Rincon-Garcia2016}, Joule heating \cite{Lee2013}, and thermal
conductance \cite{Segal2016}. In this sense, the investigation of phonon
transport in these atomic-scale junctions is presently attracting a lot of
attention for two basic reasons. On the one hand, molecular
  junctions offer the possibility to study phonon conduction in an interesting
  regime, where the system size is smaller than the phonon inelastic mean free
  path. On the other hand, the phonon contribution to the thermal conductance
plays a fundamental role, when assessing the performance of molecular junctions
as thermoelectric devices. This performance is characterized by the so-called
figure of merit $ZT$, which is given by a combination of several transport
quantities as follows \cite{Goldsmid2016}
\begin{equation}
  ZT = \frac{GS^2T}{\kappa}=\frac{Z_{\rm el}T}{1+\kappa_{\rm other}/\kappa_{\rm el}}.
  \label{eq-ZT}
\end{equation}
Here, $G$ is the electrical conductance, $S$ the thermopower, $T$ the absolute
temperature, and $\kappa$ the thermal conductance. Strictly speaking, this
thermal conductance should include all possible contributions, and it can be
written as $\kappa = \kappa_{\rm el} + \kappa_{\rm other}$, where $\kappa_{\rm el}$
is the electronic contribution and $\kappa_{\rm other}$ includes the contributions
of other heat carriers like that of phonons and photons. By bringing Eq.~(\ref{eq-ZT})
into the form on the right hand side using $Z_{\rm el}T=GS^2T/\kappa_{\rm el}$, it
is obvious that any additional heat transport contribution (beyond
electrons) will be detrimental for the thermoelectric performance, since $ZT$ needs 
to be maximized. Therefore, the experimental and theoretical determination of
$\kappa_{\rm other}$ is crucial to critically evaluate, whether molecular
junctions can potentially operate as efficient nanoscale thermoelectric devices.

The most obvious additional contribution to the heat
  conductance in molecular junctions is that of phonons, a topic that is
  currently attracting a lot of attention. For a recent review, we recommend
  Ref.~\onlinecite{Segal2016}. However, photons can also give a significant
contribution to the total heat conductance. Molecular junctions actually
constitute nanoscale gaps between metal surfaces, bridged by few or single
molecules. Thus, even if the experiments are carried out in ultra-high vacuum
conditions, one should also consider the contribution of thermal radiation or
photon tunneling. It has been understood that when two bodies are brought
sufficiently close together (with a separation below the thermal wavelength,
which is $9.6$ $\mu$m at room temperature), near-field contributions in the
form of evanescent waves dominate the radiative heat transfer and lead to a
huge enhancement of the radiative thermal conductance \cite{Polder1971}. This
near-field radiative heat transfer (NFRHT) can exceed by orders of magnitude
the limit set by the Stefan-Boltzmann law for black bodies, see
Ref.~\onlinecite{Song2015a} for a recent review. These ideas have been
experimentally verified in recent years, and advances in nanothermometry have
even made possible to explore thermal radiation in the extreme near-field
regime, where objects are separated by gaps of a few nanometers and even below
\cite{Kittel2005,Worbes2013,Kim2015,Kloppstech2017,Cui2017c}.  This aspect of
photonic heat conduction has traditionally been ignored thus far in the field
of molecular electronics.

The main goal of this work is to rigorously compute the contribution
$\kappa_{\rm pn}$ of phonons to both the thermal conductance and the figure of
merit in C$_{60}$-based single-molecule junctions using parameter-free
\emph{ab initio} electronic structure methods. But also the
  photonic part $\kappa_{\rm pt}$ will be estimated using simple models for
  the molecular junction geometries within the framework of fluctuational
  electrodynamics. We will thus ultimately consider $\kappa = \kappa_{\rm
  el} + \kappa_{\rm pn} + \kappa_{\rm pt}$ in the following, i.e., a thermal
conductance $\kappa$ consisting of electronic (el), phononic (pn), and
photonic (pt) parts.
 
The fullerene C$_{60}$ is a test-bed molecule for molecular electronics and
its electrical transport properties have been extensively investigated both
experimentally \cite{Joachim1995,Park2000,
  Parks2007,Boehler2007,Neel2007,Kiguchi2009,Schull2009,Schull2011a,Schull2011b}
and theoretically
\cite{Neel2007,Schull2009,Schull2011a,Schull2011b,Palacios2001a,Palacios2001b,Stadler2007,
  Ono2007,Shukla2008,Zheng2009,Abad2010a,Abad2010b,Bilan2012,Geranton2013}. Also
in the context of thermoelectricity, C$_{60}$-based single-molecule junctions
have been analyzed and several groups have reported room-temperature
thermopower measurements
\cite{Yee2011,Evangeli2013,Lee2014}. In particular, Evangeli
  \emph{et al.}\ \cite{Evangeli2013} employed a scanning tunneling microscope
  (STM) setup to report simultaneous measurements of the conductance and
  thermopower in ``single-C$_{60}$'' or monomer molecular junctions with Au
  electrodes. They showed that these junctions can exhibit thermopower values
ranging from $-40$ to 0 $\mu$V/K, depending of the contact details, with a
mean value of $-18$ $\mu$V/K. The findings agree well with theoretical
expectations that predict that these negative values are due to charge
transport that is dominated by the lowest unoccupied molecular orbital (LUMO)
\cite{Bilan2012}. Moreover, these authors were able to pick up C$_{60}$
molecules with the STM tip and subsequently use them to contact another
individual C$_{60}$ molecule, forming in this way molecular junctions with a
C$_{60}$ dimer bridging the gap between the Au electrodes.
These ``two-C$_{60}$'' or dimer molecular junctions were shown
  to exhibit negative thermopower values of up to $-72$ $\mu$V/K with an
  average value of $-33$ $\mu$V/K. This almost doubles the magnitude observed
  in the monomer junctions, as it is generally expected for increasing
  molecular length in the off-resonant transport regime \cite{Pauly2008a}.
These results, together with first-principles transport calculations, led the
authors to suggest that stacks of C$_{60}$ molecules could provide a way to
achieve high $ZT$ values (even above 1), making fullerene-based junctions very
promising for thermoelectric applications.  However, it is worth stressing
that this appealing suggestion was made without taking into account the phonon
contribution in the \emph{ab initio} calculations and without measuring the
thermal conductance. Indeed experimental setups were until recently not
sensitive enough to measure $\kappa$ at the atomic scale, and single-molecule
junctions have yet to be examined with newly developed tools
\cite{Cui2017a}. Thus, this interesting suggestion still has to stand careful
theoretical and experimental tests with access to all the major quantities
determining the figure of merit.

The goal of this work is to fill this gap from the theoretical point of
view. For this purpose, we present here a detailed study of
the role of the phonon transport in both the thermal conductance and the
figure of merit of monomer and dimer C$_{60}$ molecular junctions. This
study is based on a state-of-the-art combination of density functional theory
(DFT) with nonequilibrium Green's function (NEGF) techniques
\cite{Buerkle2015,Kloeckner2016}, which allows us to compute the contribution
of both electrons and phonons to the different transport properties. Our results 
show that the phonons play a minor role in the thermal conductance of the monomer 
junctions, while they largely determine this property in dimer junctions. This fact 
results in a substantial reduction of the $ZT$ values of the dimer junctions, as
compared to the monomer junctions, in spite of the fact that phonons are transported
less efficiently in the dimer case. In other words, our results do not back
up the suggestion above, but instead they show that phonons severely limit
the thermoelectric performance of dimer junctions. In addition, we provide
in this work a critical analysis of the impact of thermal radiation on the
$ZT$ values of molecular junctions, a factor that has been overlooked so far
in molecular electronics. We show, in particular, that the NFRHT between the
metallic electrodes can indeed further reduce the figure of merit of molecular
junctions. This effect is particularly pronounced in the tunneling
regime. Overall, our work demonstrates the importance of taking into account
both phonons and photons for a proper evaluation of the performance of
molecular junctions as possible thermoelectric
devices. Moreover, it provides valuable insights into the
  relative contribution of different heat carriers to the thermal conductance,
  namely electrons, phonons, and photons.

The rest of the paper is organized as follows. First, in
section~\ref{sec-Methods} we briefly describe the theoretical methods employed
to obtain the results presented in this work. Then, in
section~\ref{sec-Results} we discuss the main results, concerning in
subsection~\ref{subsec-Results-epn} the impact of phonons on the thermal
conductance and figure of merit of C$_{60}$-based molecular junctions and 
in subsection~\ref{subsec-Results-pt} the importance of near-field
thermal radiation. Finally, we summarize in section~\ref{sec-Conclusions} our
main conclusions.

\section{Theoretical approach} \label{sec-Methods}

Our primary goal is to compute the different charge and energy transport
properties that determine the figure of merit of single-molecule
junctions. Electronic and phononic ones are treated within the
  framework of the Landauer-B\"uttiker approach to coherent transport. For
this purpose, we make use of the first-principles formalism developed by us
and reported in
Refs.~\onlinecite{Pauly2008,Buerkle2015,Kloeckner2016}. Our
  approach is based on a combination of DFT and NEGF techniques and allows us
  to compute all of the basic thermoelectric linear response transport
  properties of a nanoscale system, namely $G$, $S$, $\kappa_{\rm el}$ and
  $\kappa_{\rm pn}$.  Photon transport, on the other hand, is described
within the framework of fluctuational electrodynamics. In particular, we
compute $\kappa_{\rm pt}$ for nanometer-sized gaps following
Ref.~\onlinecite{Song2015b}. In what follows, we briefly describe the main
features of our methods and refer to the above-mentioned references for
further details.

\subsection{Contact geometries, electronic structure, and vibrational properties}

Our modeling starts with the construction of the molecular junction
geometries. We use DFT to obtain equilibrium geometries through total energy
minimization and to compute their electronic structure. Vibrational properties
of the optimized equilibrium contacts are subsequently obtained in the
framework of density functional perturbation theory. For these purposes, we
use procedures as implemented in the quantum chemistry software package
TURBOMOLE 6.5 \cite{TURBOMOLE,Deglmann2002,Deglmann2004}. In our DFT
calculations we employ the Perdew-Burke-Ernzerhof exchange-correlation
functional \cite{Perdew1992,Perdew1996} with the Grimme dispersion correction
\cite{Grimme2010}, the “default2” basis set of split-valence-plus-polarization
quality def2-SV(P) \cite{Weigend2005}, and the corresponding Coulomb fitting
basis \cite{Weigend2006}. In order to accurately determine the vibrational
energies and force constants, we use very strict convergence criteria. In
particular, total energies are converged to a precision of better than
$10^{-9}$~a.u., whereas geometry optimizations are performed until the change
of the maximum norm of the Cartesian gradient is below $10^{-5}$~a.u..

\subsection{Electronic transport}\label{subsec-theory-el}

Within the Landauer-B\"uttiker picture, the contribution of electrons to the
different transport properties is determined by the energy-dependent
electronic transmission $\tau_{\rm el}(E)$. In particular, in the linear
response regime, in which we are interested in, the electrical conductance
$G$, thermopower $S$, and the electronic thermal conductance
$\kappa_{\textnormal{el}}$ are given by \cite{Sivan1986,Cuevas2010}
\begin{eqnarray}
G & = & G_0 K_{0}, \label{eq-G}\\ 
S & = &-\frac{K_{1}}{eTK_{0}}, \label{eq-S} \\ 
\kappa_{\mathrm{el}} & = & \frac{2}{hT} \left(K_{2}-\frac{K_{1}^{2}}{K_{0}}\right),
\label{eq-kel}
\end{eqnarray}
where $e=\left|e\right|$ is the absolute value of the electron charge, $h$ is
Planck's constant, $k_{\rm B}$ is Boltzmann's constant, $T$ is the average
junction temperature, and $G_0 = 2e^2/h$ is the conductance quantum. The
coefficients $K_n$ in Eqs.~(\ref{eq-G})-(\ref{eq-kel}) are given by
\begin{equation}
  K_n = \int^{\infty}_{-\infty} dE \, \tau_{\rm el}(E) \left(-\tfrac{\partial
    f(E)}{\partial E}\right)(E-\mu)^{n},
  \label{eq-Kn}
\end{equation}
where $f(E) = \left\{ \exp[(E-\mu)/k_{\mathrm{B}}T]+1\right\} ^{-1}$ is the
Fermi function. Here, the chemical potential $\mu\approx E_{\textnormal{F}}$
is approximately given by the Fermi energy $E_{\textnormal{F}}$ of the Au
electrodes. Dependences of transport quantities, the coefficients $K_n$ and
the Fermi function on temperature and chemical potential have been suppressed.

Let us emphasize that we have used the exact Eqs.~(\ref{eq-G})-(\ref{eq-kel})
in our calculations, but it is instructive to have in mind the corresponding
low-temperature expansions, which turn out to be very good approximations in
almost all cases. They read
\begin{eqnarray}
G & \approx & G_0 \tau_{\rm el}(E_{\rm F}), \label{eq-G-lowT}\\
S & \approx & - \frac{\pi^2 k^2_{\rm B} T}{3e} \left. \frac{\partial_E \tau_{\rm el}(E)}
{\tau_{\rm el}(E)}\right|_{E=E_{\rm F}}, \label{eq-S-lowT} \\
\kappa_{\mathrm{el}} & \approx & L_0 G T.
\label{eq-kel-lowT}
\end{eqnarray}
The latter expression for $\kappa_{\mathrm{el}}$ is known as the
Wiedemann-Franz law \cite{Cuevas2010} and $L_0 = (k_{\rm B}/e)^2 \pi^2/3$ is
the Lorentz number.

We have computed the electron transmission by making use of our DFT-NEGF
formalism, implemented in TURBOMOLE and explained in detail in
Ref.~\onlinecite{Pauly2008}. In particular, to construct the electrode surface
Green's function, we use a broadening of
$\eta=10^{-2}$~a.u.\ and $32\times32$ $k$-points in the
transverse direction, which was found to be sufficient to converge the
electronic transport coefficients.

\subsection{Phonon transport}\label{subsec-theory-pn}

In analogy with the electronic part, we have computed the phononic
contribution to the heat conductance within the framework of the
Landauer-B\"uttiker picture. By choosing this procedure, we ignore anharmonic
effects that are expected to play a minor role in short molecular junctions
like the ones studied here \cite{Segal2016}. Within this picture, the phonon
contribution to the heat conductance in the linear regime can be expressed in
terms of the phononic transmission function $\tau_{\rm pn}(E)$ as follows
\cite{Rego1998,Mingo2003}
\begin{equation}
  \kappa_{\rm pn} = \frac{1}{h} \int_{0}^{\infty} dE \,E \tau_{\rm pn}(E)
  \frac{\partial n(E,T)}{\partial T}, \label{eq-kph}
\end{equation}
where $n(E,T)=[\exp(E/k_{\rm B}T)-1]^{-1}$ is the Bose function. As for the
electronic transport quantities, we have suppressed the temperature dependence
of $\kappa_{\rm pn}$.

For computing the phononic thermal conductance, we need to determine the
energy-dependent phonon transmission $\tau_{\rm pn}(E)$. To do so, we have
employed the procedures described in
Refs.~\onlinecite{Buerkle2015,Kloeckner2016}.  A broadening of
$\eta=10^{-5}$~a.u.\ and $32\times32$ $k$-points in the
transverse direction were used to obtain well-converged phononic transport
properties. 

\subsection{Photon transport}\label{subsec-theory-pt}

We use fluctuational electrodynamics \cite{Rytov1989} to determine
$\kappa_{\rm pt}$. This formulation of NFRHT was indeed employed recently to
study the radiative heat transfer between a gold surface and a gold tip in the
extreme near-field regime and shown to work nicely all the way down to gaps of
a few nanometers \cite{Kim2015}.

Our calculation of the radiative thermal conductance
  $\kappa_{\rm pt}$ proceeds in two steps. First, we calculate the so-called
  heat transfer coefficient, i.e.\ the linear radiative thermal conductance
  per unit area, for two Au infinite parallel plates, and then we use this
  result together with the so-called proximity approximation (see below) to
  compute $\kappa_{\rm pt}$ for two different geometries (see
  Fig.~\ref{fig-RHT}(a)): (i) a tip-surface geometry and (ii) a tip-tip
  geometry.

Within the framework of fluctuational electrodynamics \cite{Rytov1989} the
heat transfer coefficient $\zeta$ for two infinite parallel plates separated
by a distance $\Delta$ is given by \cite{Polder1971}
\begin{equation}
  \label{eq-HTC}
  \zeta(\Delta) = \int^{\infty}_{0} \frac{d \omega}{2\pi} \frac{\partial
    \Theta(\omega,T)}{\partial T} \int \frac{d^2k_\parallel}{(2\pi)^2}
  \tau_{\rm pt}(\omega,\boldsymbol{k}_\parallel) ,
\end{equation}
where $\Theta(\omega,T) = \hbar \omega n(E,T)$, $\omega$ is the radiation
frequency, $\boldsymbol{k} = (k_\bot,\boldsymbol{k}_\parallel)$ is the
wavevector expressed in terms of components perpendicular and parallel to the
surface planes with $k_\bot = k_{\rm x}$ and $\boldsymbol{k}_\parallel =
(k_{\rm y},k_{\rm z})$, $\tau_{\rm pt}(\omega,\boldsymbol{k}_\parallel)$ is
the total transmission probability of the electromagnetic waves, and we have
omitted the temperature dependence of $\zeta$. Notice that the second integral
in Eq.~(\ref{eq-HTC}) is carried out over all possible directions of
$\boldsymbol{k}_\parallel$, and it includes the contribution of both
propagating waves with $k_\parallel < \omega/c$ and evanescent waves with
$k_\parallel > \omega/c$, where $k_\parallel = |\boldsymbol{k}_\parallel|$ and
$c$ is the velocity of light in vacuum. The total transmission can be written
as $\tau_{\rm pt}(\omega,\boldsymbol{k}_\parallel) =
\tau_{\rm s}(\omega,\boldsymbol{k}_\parallel) +
\tau_{\rm p}(\omega,\boldsymbol{k}_\parallel)$, where the contributions of s- and
p-polarized waves are given by \cite{Polder1971}
\begin{eqnarray}
  \label{eq-trans-photon}
  \tau_{\alpha}(\omega,\boldsymbol{k}_\parallel) = \hspace{6cm} && \\ \left\{ \begin{array}{ll}
    (1 - |r_{\alpha,21}|^2) ( 1 - |r_{\alpha,23}|^2)/|D_{\alpha}|^2 , \;\; k_\parallel < \omega/c \\
    4 \mbox{Im} \{ r_{\alpha,21} \} \mbox{Im} \{ r_{\alpha,23} \} e^{-2|k_{\bot,2}|\Delta} /|D_{\alpha}|^2
    , \;\; k_\parallel > \omega/c \end{array} \right. . && \nonumber
\end{eqnarray}
Here, $\alpha={\rm s,p}$ and index 1 refers to the left
plate, 2 to the vacuum gap, and 3 to the right plate. The coefficients
$r_{\alpha,ij}$ are reflection or Fresnel coefficients of the
two interfaces between gold and the vacuum gap and are given by
\begin{eqnarray}
\label{eq-rs}
r_{{\rm s},ij} = \frac{k_{\bot,i} - k_{\bot,j}}{k_{\bot,i} + k_{\bot,j}} \;\; \mbox{and} \; \; 
r_{{\rm p},ij} = \frac{\epsilon_j k_{\bot,i} - \epsilon_i k_{\bot,j}}{\epsilon_j k_{\bot,i} + \epsilon_i k_{\bot,j}} ,
\end{eqnarray}
where the component of the wavevector in system $i$
perpendicular to the plates may also be expressed as $k_{\bot,i} =
\sqrt{\epsilon_i \omega^2 /c^2 - k_\parallel^2}$ and $\epsilon_i(\omega)$ is
the corresponding dielectric function. Finally, $D_{\alpha} = 1 -
r_{\alpha,21} r_{\alpha,23} e^{2ik_{\bot,2}\Delta}$ is a Fabry-P\'erot-like
denominator, resulting from the multiple scattering between the two
interfaces.

To compute the heat transfer coefficient, we have employed the experimental
dielectric function for Au reported in Ref.~\onlinecite{Ordal1983}. Our
results basically coincide with those reported in
Ref.~\onlinecite{Chapuis2008}, with minor differences due to the different Au
dielectric function employed here. We also find for small gaps in the
near-field regime that the contribution of s-polarized evanescent waves,
resulting from total internal reflection, completely dominates the radiative
heat transfer all the way down to separations of about 1~\AA. Let us remark
that in the formalism detailed above, we make use of a local approximation, in
which the dielectric function is assumed to depend only on frequency. However,
non-local contributions due to the momentum dependence of the dielectric
function have been shown to be negligible for gaps larger
than 1~\AA{} \cite{Chapuis2008}, as the ones studied in this
work.

We can use the results for the heat transfer coefficient $\zeta$ to estimate
the radiative thermal conductance in a junction with Au electrodes. For this
purpose, we need to know the macroscopic shape of the electrodes. In STM-based
experiments, the electrodes are a tip and planar surface. Thus, and since we
are interested in the extreme near-field regime (with gaps in the order of
nanometers), it is reasonable to model this situation with a finite sphere of
a given radius $R$ and an infinite planar surface. We shall refer to this
geometry as tip-surface geometry. On the other hand, in the case of
mechanically controllable break junctions, it is more appropriate to model the
electrodes as two spherical tips. For simplicity we assume two spheres of the
same radius $R$. We shall refer to this geometry as tip-tip geometry.

In principle, one can carry out a very accurate analysis of the radiative heat
transfer in these two types of geometries along the lines of
Ref.~\onlinecite{Kim2015}, but for our purposes here it suffices to make use of the
so-called proximity approximation, sometimes referred to as Derjaguin
approximation \cite{Derjaguin1956}. It has been shown to provide a very good
approximation for the two geometries considered here in the limit in which the
tip radius is much larger than the gap size
\cite{Sasihithlu2011,Otey2011,Song2015b,Kim2015}. In particular, Kim \emph{et
al.}\ \cite{Kim2015} showed that microscopic details like surface roughness
(either at the tip or at the surface) do not significantly change the results
for Au. Within the proximity approximation the radiative heat conductance
between a sphere and a plane and between two spheres can be computed as
\begin{equation}
  \label{eq-PA}
  \kappa_{\rm pt}(\Delta) = \int^R_0 \zeta(h(r)) 2\pi r \, dr ,
\end{equation}
where $\Delta$ is the gap, $\zeta$ is the heat transfer coefficient calculated
as described above, $R$ is the sphere radius, and $h(r)= \Delta + R -
\sqrt{R^2 - r^2}$ for the tip-surface geometry and $h(r)= \Delta + 2R -
\sqrt{R^2 - r^2}$ for the tip-tip geometry, as depicted in
Fig.~\ref{fig-RHT}(a).

  Finally, let us point out that we do not take the presence of
  molecules into account in the calculation of the photonic thermal
  conductance. In the single-molecule junctions considered in this work, the
  molecules only modify the refractive index of the gap in a very tiny
  region. Due to the long-wavelength nature of the electromagnetic waves that
  dominate the NFRHT, this region is orders of magnitude smaller than the
  portion of the electrodes that contribute to the radiative heat transfer
  \cite{Kim2015}. Therefore, the role of the molecules in the photonic
  transport is expected to be negligible.

\section{Results}\label{sec-Results}

\subsection{Charge transport, phonon transport and thermoelectric
  figure of merit}\label{subsec-Results-epn}

\begin{figure}[t]
\begin{center} \includegraphics[width=\columnwidth,clip]{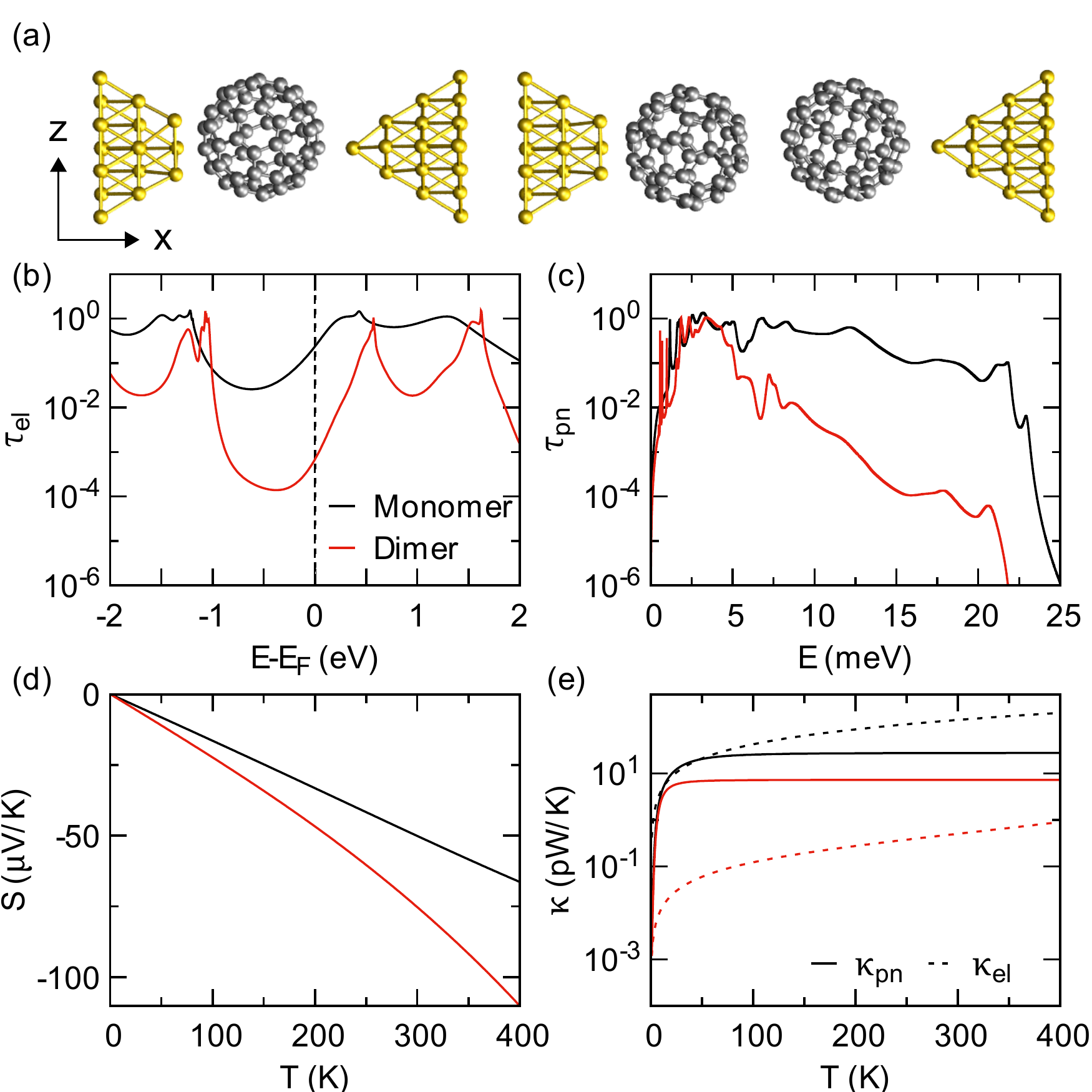} \end{center}
\caption{(Color online) (a) Equilibrium geometries of C$_{60}$ monomer and
  C$_{60}$ dimer junctions. The molecules are bonded to blunt
    and sharp Au electrodes on the left and right. These geometries correspond
    to the minimum of the total energy with respect to the distance $d$
    between the electrodes. We call this particular separation for the
    equilibrium geometry $d_0$, see illustration in Fig.~\ref{fig-Taud}(a).
  (b) Electronic transmission as a function of energy, measured with respect
  to the Fermi level, for the two geometries shown in panel (a). (c) Phonon
  transmissions as a function of energy. (d) Thermopower of the two junctions
  as a function of temperature. (e) Electronic and phononic contributions to
  the heat conductance as a function of temperature for both junctions. Data
  referring to the monomer is shown in black, those of the dimer in red.}
\label{fig-d0}
\end{figure}

\begin{figure}[t!]
\begin{center} \includegraphics[width=\columnwidth,clip]{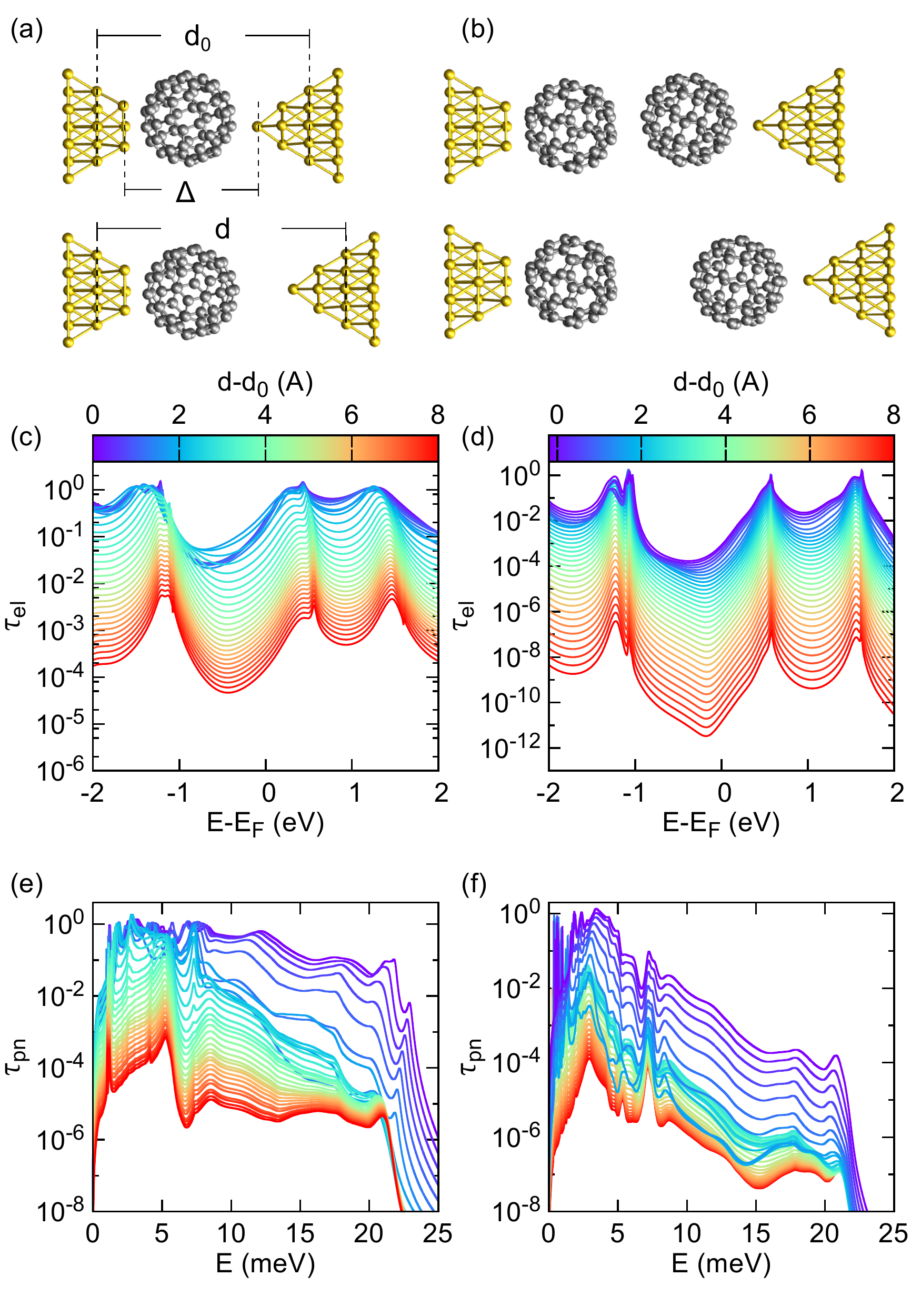} \end{center}
\caption{(Color online) Evolution of the electronic and phononic transmissions
  with the elongation of the junctions in Fig.~\ref{fig-d0}. (a) Geometries of
  the monomer junction. The upper geometry is the equilibrium geometry,
  already displayed in Fig.~\ref{fig-d0}, while the lower junction is
  stretched. The distance $d$ is measured between Au layers of the electrodes
  that are held fixed in the geometry optimization process, and $d_0$ is the
  separation for the equilibrium geometry. The gap distance $\Delta$ between
  the outer atoms of the Au electrodes that contact the molecules is somewhat
  shorter. (b) Same as in (a) but for the dimer junction. (c) Evolution of the
  electronic transmission of the monomer contact upon stretching as a function
  of the energy that is measured with respect to the Fermi energy. (d) The
  same as in panel (c) but for the dimer junction. (e) The corresponding
  evolution of the phononic transmission for the monomer junction. (f) The
  same as in panel (e) but for the dimer junction. We determine elongations as
  $d-d_0$, and plot the corresponding energy-dependent transmission curves in
  different colors as indicated by the color scale bar.}
\label{fig-Taud}
\end{figure}

We start our discussion of the results by considering the
  geometries shown in Fig.~\ref{fig-d0}(a) for an Au-C$_{60}$-Au and an
  Au-C$_{60}$-C$_{60}$-Au junction, which will be hereafter referred to as
  monomer and dimer junctions, respectively. In these particular geometries,
the molecules make contact to three Au atoms of a blunt tip on the left, while
they are bonded to a single Au atom of an atomically sharp tip at the right Au
electrode. This asymmetric situation is meant to mimic the geometries realized
in the STM experiments of Ref.~\onlinecite{Evangeli2013}, in which one of the
electrodes is an Au surface and the other one an Au STM
tip. Let us stress that these geometries were obtained by
  minimizing the total energy of the junctions as a function of the electrode
  separation $d$, yielding the distance $d_0$. The distances $d$ and $d_0$ are
  also visualized in Fig.~\ref{fig-Taud}(a). Using a coordinate system,
oriented as depicted in Fig.~\ref{fig-d0}, we find that the closest Au-C
separation along the $x$-axis is roughly 1.8~\AA{} from the blunt tip and
2.2~\AA{} from the sharp tip. The $x$-axis separation between the two C$_{60}$
molecules in the dimer junction is approximately 3.0~\AA{}. In
Fig.~\ref{fig-d0}(b)-(e) we summarize the thermoelectric transport properties
for these two junctions by showing the electronic transmission as a function
of energy, the corresponding phononic transmission, the thermopower as a
function of temperature, and the electronic and phononic contributions to the
heat conductance as a function of temperature.

For these blunt-sharp junctions, the electrical conductance is $0.23G_0$ for
the monomer and $7.1 \times 10^{-4}G_0$ for the dimer. These values are close
to the experimental values found in Ref.~\onlinecite{Evangeli2013}, where mean
values of $0.1G_0$ and $1.8 \times10^{-3}G_0$ were reported for the monomer
and dimer junctions, respectively. Notice that in both cases the electronic
transmission at the Fermi energy is determined by the LUMO of the molecules,
as it has been reported in numerous studies, see for instance
Ref.~\onlinecite{Bilan2012} and references therein.

With respect to the phononic transmission, one can see in Fig.~\ref{fig-d0}(c)
that the phonon conduction is dominated by low-lying vibrational modes with
energies $E<10$~meV. Let us recall that the Debye energy of the metal
electrodes sets an upper limit for the energy of the vibrational modes that
can contribute to the transport, which in our case is around 20~meV
\cite{Buerkle2015}. However, in the range between 10 and 20~meV there are no 
significant contributions to the phonon thermal conductance, which we attribute 
to the weak metal-molecule coupling for the modes in this energy range.

Turning now to the thermopower, we see in Fig.~\ref{fig-d0}(d) that it
approximately follows a linear dependence on temperature, as expected from the
low-temperature expression in Eq.~(\ref{eq-S-lowT}). In
particular, the room-temperature thermopower has a value of $-49.6$ $\mu$V/K
for the monomer and $-75.3$ $\mu$V/K for the dimer junction. The negative
values are due to the fact that the electronic transport is dominated in both
cases by the LUMO, meaning that electric conduction is electron-like. Notice
also that the thermopower value for the dimer junction almost doubles that of
the monomer, similar to what was observed in Ref.~\onlinecite{Evangeli2013},
while the absolute values are somewhat larger than those reported
experimentally.

The temperature dependence of the different contributions to the heat
conductance, displayed in Fig.~\ref{fig-d0}(e), shows that both junctions
behave qualitatively different, especially at room temperature. In the monomer
junction the room-temperature electronic thermal conductance $\kappa_{\rm el}
= 142.0$~pW/K, which is very close to the value expected from the
Wiedemann-Franz law, dominates over the phononic one,
$\kappa_{\rm pn} = 25.5$~pW/K. At
the contrary, for the dimer junction the phononic contribution $\kappa_{\rm
  pn} = 7.3$~pW/K dominates the thermal conductance and the electrons give an
insignificant contribution of $\kappa_{\rm el} = 0.5$~pW/K. Notice also that
the total thermal conductance at room temperature is about 20 times larger for
the monomer case than for the dimer case, which is mainly due to a dramatic
decrease in the electronic contribution for the latter. For the sake of
comparison, it is worth mentioning that we found in
Ref.~\onlinecite{Kloeckner2016} that alkane-based chains of varying length
exhibit phononic thermal conductance values ranging from 15 to 45~pW/K. Our
results for these blunt-sharp C$_{60}$ junctions show that there is a change
in the dominant heat carriers as a function of the number of C$_{60}$
molecules. This has a crucial impact on the figure of merit, as we proceed to
discuss.

\begin{figure}[bt]
\begin{center} \includegraphics[width=0.8\columnwidth,clip]{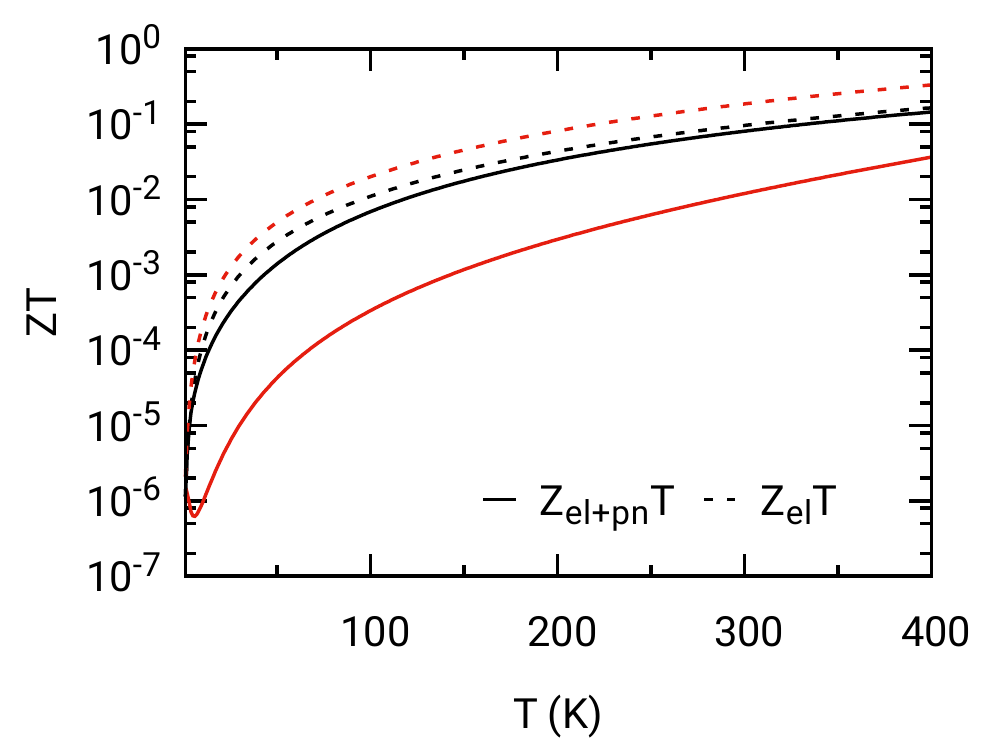} \end{center}
\caption{(Color online) Figure of merit $ZT$ as a function of temperature for
  the two geometries shown in Fig.~\ref{fig-d0}(a). Black
    curves are for the monomer junction, while red ones are for the dimer. The
    solid lines correspond to $Z_{\rm el+pn}T$, including both the electronic
    and phononic contributions to the heat conductance, $\kappa=\kappa_{\rm
      el}+\kappa_{\rm pn}$, while the dashed lines correspond to $Z_{\rm
      el}T$, including only the electronic one, $\kappa=\kappa_{\rm el}$.}
\label{fig-ZT}
\end{figure}
%

Results for the figure of merit $ZT$, see Eq.~(\ref{eq-ZT}), are shown in
Fig.~\ref{fig-ZT}.  There we depict the evolution of $Z_{\rm el+pn}T$ with
temperature, taking into account both the electronic and phononic
contributions to the thermal conductance, and $Z_{\rm el}T$, taking into
account only the electronic contribution. The first thing to notice is that at
room temperature the monomer junction reaches a value of $Z_{\rm el}T =
0.093$, which is only slightly reduced to $Z_{\rm el+pn}T = 0.079$ by the
phononic contribution to the thermal transport. However, in the dimer case the
room-temperature value $Z_{\rm el}T = 0.18$ is strongly reduced by the phonon
heat conduction to $Z_{\rm el+pn}T = 0.012$. Exploiting the Wiedemann-Franz
law in Eq.~(\ref{eq-kel-lowT}), which we find to be well obeyed, we can write
$ZT\approx S^2/[L_0(1+\kappa_{\rm other}/\kappa_{\rm el})]$. With this
relation the $ZT$ values of the dimer junction can be interpreted as
follows. Neglecting $\kappa_{\rm other}$, the increase in $Z_{\rm el}T$ for
the dimer junction is due to the increase of the thermopower as compared with
the monomer. Setting $\kappa_{\rm other}=\kappa_{\rm pn}$, we
  see from Fig.~\ref{fig-d0} that the phonon thermal conductance of the dimer
  is slightly lower than those of the monomer. But due to the much stronger
  decrease of the electrical conductance for two C$_{60}$ as compared to a
  single one, leading to a corresponding reduction of $\kappa_{\rm el}$ via
  the Wiedemann-Franz law, the ratio $\kappa_{\rm pn}/\kappa_{\rm el}$ and the
  denominator $L_0(1+\kappa_{\rm pn}/\kappa_{\rm el})$ get large, resulting in
  a pronounced suppression of $Z_{\rm el+pn}T$ for the dimer. Thus, this
first example of blunt-sharp monomer and dimer junctions indicates that
because of the important contribution of phonon conduction to the thermal
transport, the appealing suggestion of Ref.~\onlinecite{Evangeli2013} that
high $ZT$ values may be achieved by stacking C$_{60}$ molecules, as discussed
in the introduction, is not backed up by our calculations.

\begin{figure}[t]
  \begin{center} \includegraphics[width=\columnwidth,clip]{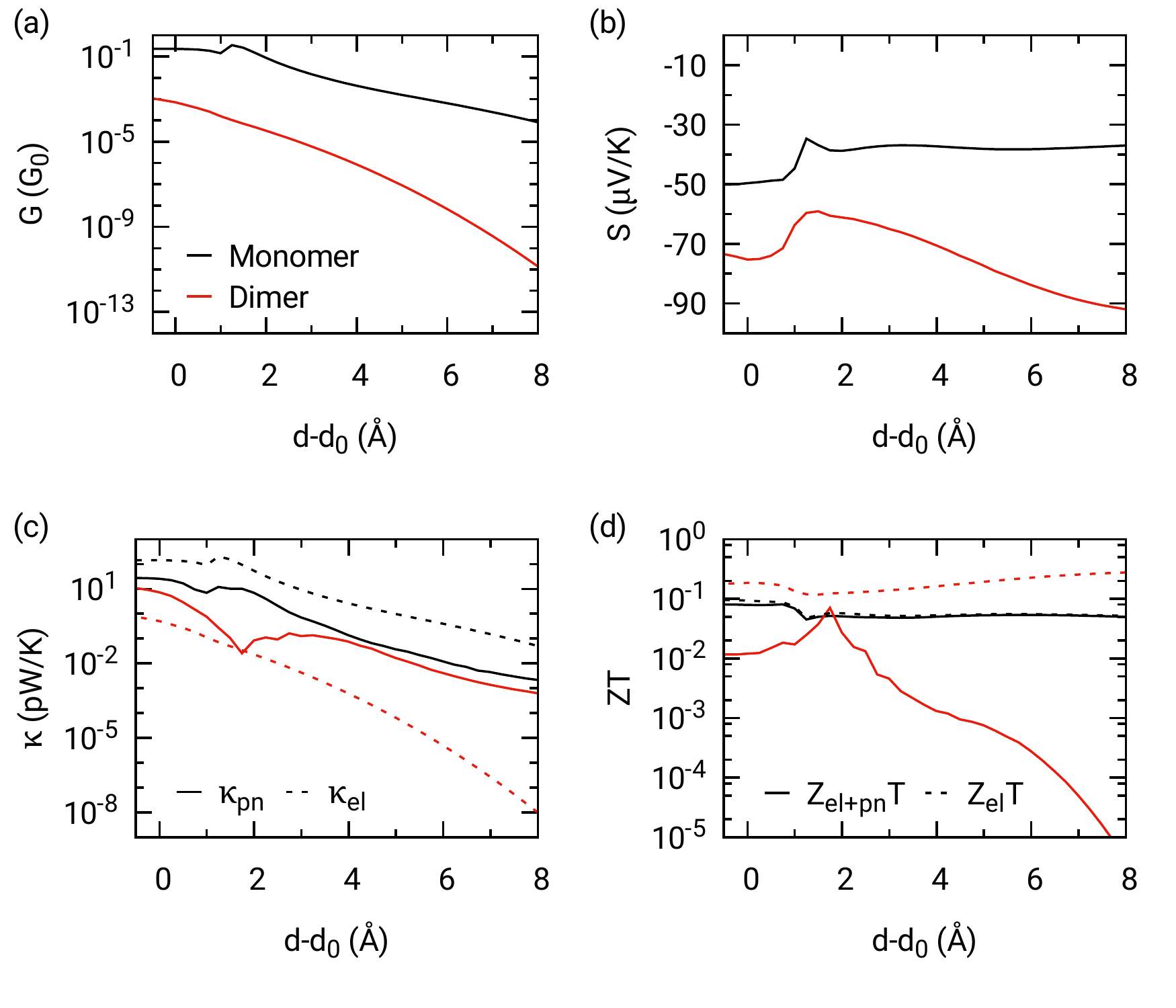} \end{center}
  \caption{(Color online) Transport coefficients at $T=300$~K as a function of
    the distance $d-d_0$. The results are obtained from the transmission
    functions shown in Fig.~\ref{fig-Taud}. (a) Electrical conductance, (b)
    thermopower, (c) electronic and phononic thermal conductance, and (d) the
    corresponding figures of merit $Z_{\rm el+pn}T$ and $Z_{\rm el}T$. In all
    panels, black lines correspond to the monomer junction and red lines to
    the dimer junction.}
  \label{fig-Resultd}
\end{figure}

To test the robustness of the main conclusions so far, especially the strong
reduction of $Z_{\rm el+pn}T$ as compared to $Z_{\rm el}T$ for the dimer
junction, we have studied the role of strain in the different transport
properties. For this purpose, and in order to mimic the STM experiments of
Ref.~\onlinecite{Evangeli2013}, we simulated the stretching and compression of
the junctions shown in Fig.~\ref{fig-d0}. The evolutions of the electronic and
phononic transmissions for the monomer and dimer junctions are shown in
Fig.~\ref{fig-Taud}. Here, the different curves correspond to different
distances $d$ between the electrodes, measured with respect to the distance in
the equilibrium geometries $d_0$, as defined in Fig.~\ref{fig-Taud}(a). The
thermoelectric transport properties $G$, $S$, $\kappa$ and $ZT$ are shown in
Fig.~\ref{fig-Resultd} as a function of $d-d_0$. As one can see in
Fig.~\ref{fig-Resultd}(a), the electrical conductance exhibits an exponential
decay both for the monomer and the dimer, as expected, when the junctions
break and enter the tunneling regime. At all distances, $G$ is
  much lower for the dimer than for the monomer, as expected for off-resonant
  transport (see for instance the electronic transmissions in
  Fig.~\ref{fig-Taud}(c,d)). In this regime the thermopower remains
relatively constant, see Fig.~\ref{fig-Resultd}(b), and for any distance the
value for the dimer junction roughly doubles that of the monomer junction.
With respect to the thermal conductance, both electronic and phononic
contributions decay monotonically in the tunneling regime. And, like in the
equilibrium geometry, we find that while for the monomer the electrons
determine the thermal transport, in the dimer case the phonons dominate at
almost all distances due to the largely reduced electronic
  thermal conductance of the dimer junctions. This fact is reflected in the
behavior of $Z_{\rm el+pn}T$ for both junctions, see
Fig.~\ref{fig-Resultd}(d), which is dictated by the electronic thermal
conductance in the monomer case during the whole stretching process, while it
is clearly limited by the phononic thermal conductance in the dimer
junction. In other words, the main conclusion of the previous paragraph --
that $ZT$ values of dimer junctions are small due to phonon heat conduction --
holds for the whole elongation process from the contact to the deep tunneling
regime.

\begin{table*}[t]
\caption{\label{table1} Gap between the gold electrodes
    $\Delta$ and room temperature values of the different transport properties
    for the three types of monomer and dimer equilibrium geometries with
    $d=d_0$ of Figs.~\ref{fig-d0}, \ref{fig-TopTop}, \ref{fig-HolHol}. The
    distances $d,d_0,\Delta$ are defined in Fig.~\ref{fig-Taud}.}
\begin{tabular}{cccccccc}
{\bf junction type} \; & \; $\Delta$ (nm) \; & \; $G$ ($G_0$) \; & \; $S$
($\mu$V/K) \; & \; $\kappa_{\rm el}$ (pW/K) \; & \;
$\kappa_{\rm ph}$ (pW/K) \; & \; $Z_{\rm el}T$ \; & \; $Z_{\rm el+pn}T$ \; \\ \hline monomer,
blunt-sharp tips & 1.12 & 0.23 & -49.6 & 142.0 & 25.5 & 0.093 & 0.079
\\ dimer, blunt-sharp tips & 2.12 & $7.1 \times 10^{-4}$ & -75.3 & 0.5 &
 7.3 & 0.185 & 0.012 \\ monomer, sharp-sharp tips & 1.18 & 0.11 & -57.4 & 68.6
& 20.0 & 0.129 & 0.100 \\ dimer, sharp-sharp tips & 2.18 & $1.3 \times
10^{-4}$ & -189.2 & 0.1 & 7.4 & 0.963 & 0.014 \\ monomer, blunt-blunt tips &
1.11 & 1.08 & -41.9 & 572.7 & 46.3 & 0.077 & 0.071 \\ dimer, blunt-blunt
tips & 1.91 & $2.3 \times 10^{-3}$ & -96.9 & 1.8 & 7.0 & 0.273 & 0.057
\\ \hline
\end{tabular}
\end{table*}

\begin{figure}[b]
\begin{center} \includegraphics[width=\columnwidth,clip]{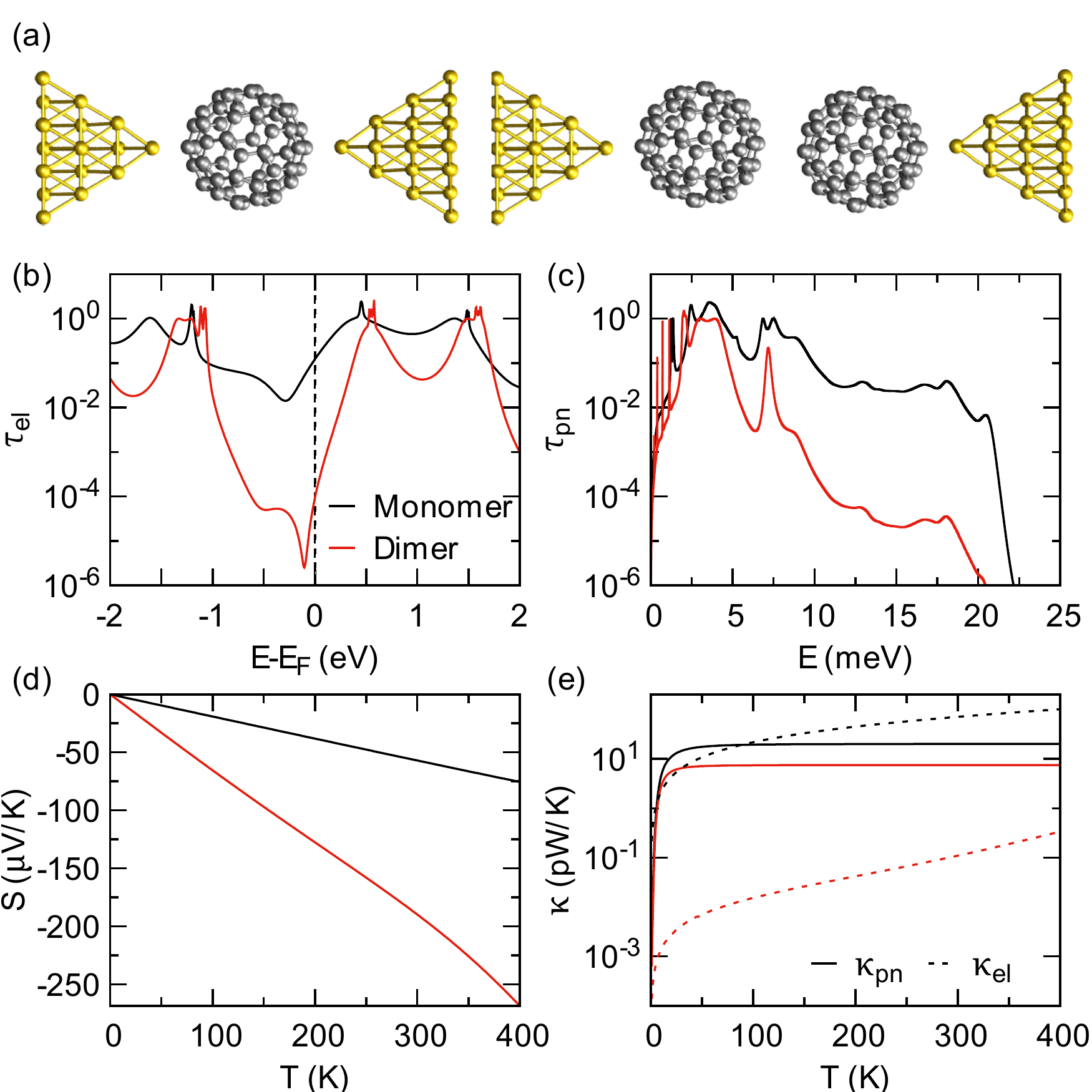} \end{center}
\caption{(Color online) (a) Equilibrium geometries of a C$_{60}$ monomer and a
  C$_{60}$ dimer, bonded to both Au electrodes through atomically sharp tips.
  These geometries correspond to the minimum of the total
    energy with respect to the distance $d$ between the electrodes,
    i.e.\ $d=d_0$. (b) Electronic transmission as a function of energy,
  measured with respect to the Fermi level, for the two geometries shown in
  panel (a). (c) Phonon transmissions as a function of energy. (d) Thermopower
  of the two junctions as a function of temperature. (e) Electronic and
  phononic contributions to the heat conductance for monomer and dimer
  junctions as a function of temperature. Data referring to the monomer is
  shown in black, those of the dimer in red.}
\label{fig-TopTop}
\end{figure}

For completeness, let us now consider the role of the binding geometry. In
particular, we now proceed to discuss the results for the equilibrium
geometries shown in Fig.~\ref{fig-TopTop}(a), where the molecules are bonded
to both Au electrodes through atomically sharp tips. This kind of symmetric
geometry is more likely to be realized in a break-junction configuration like
that of Ref.~\onlinecite{Boehler2007}. The results for the different transport
properties $G$, $S$, $\kappa_{\rm el}$ and $\kappa_{\rm pn}$ are qualitatively
similar to those of the blunt-sharp geometries discussed above. All the room
temperature values of these properties are summarized in Table~\ref{table1}
together with the derived $Z_{\rm el}T$ and $Z_{\rm el+pn}T$ values. The main
difference to the blunt-sharp geometries is the fact that the electronic
transmission does not follow a Lorentzian-like shape around the Fermi energy,
see Fig.~\ref{fig-TopTop}(b). This peculiar energy dependence is the result of
a quantum interference and its origin has been explained in detail in
Ref.~\onlinecite{Geranton2013}. This quantum interference leads to a rather
steep transmission at the Fermi energy, which is in turn responsible for the
particularly large values of the thermopower for these kind of binding
geometries, see Fig.~\ref{fig-TopTop}(d). The phononic thermal
  conductance in Fig.~\ref{fig-TopTop}(e) is again determined by the
  transmission at energies below 10~meV, as visible in
  Fig.~\ref{fig-TopTop}(c). Finally, $\kappa_{\rm el}$ in
  Fig.~\ref{fig-TopTop}(e) is larger than $\kappa_{\rm pn}$ at $T=300$~K for
  the monomer junction. This behavior is reversed for the dimer junction,
where $\kappa_{\rm pn} \gg \kappa_{\rm el}$.

\begin{figure}[b]
  \begin{center} \includegraphics[width=0.8\columnwidth,clip]{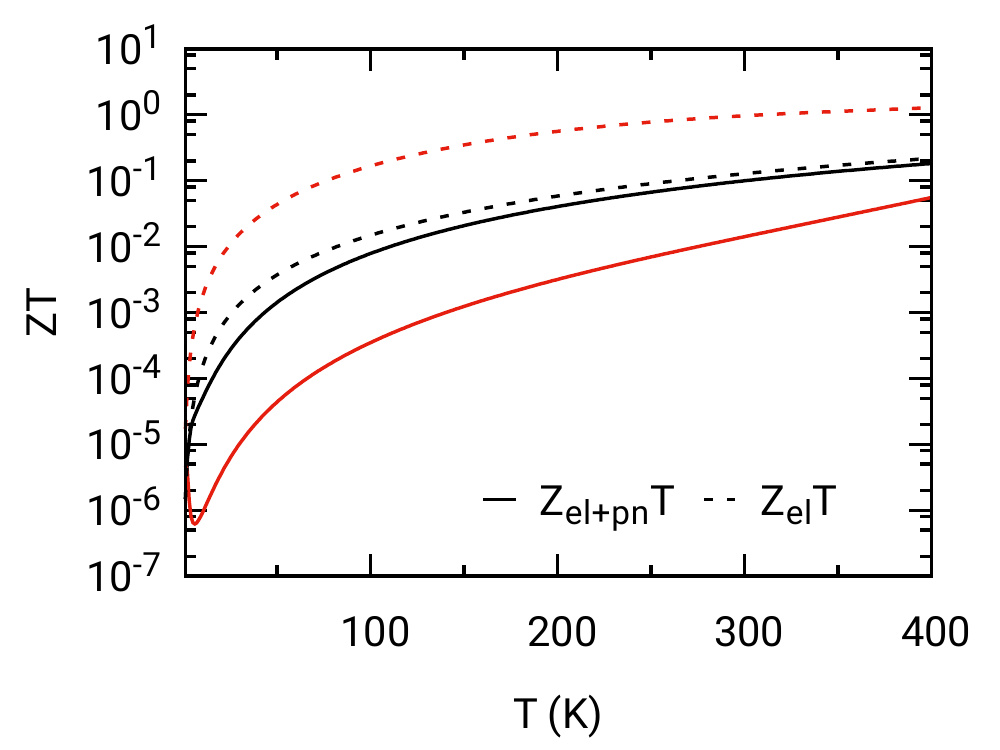} \end{center}
  \caption{(Color online) Figure of merit $ZT$ as a function of temperature
    for the two geometries shown in Fig.~\ref{fig-TopTop}(a). Black curves are
    for the monomer junction, while red ones are for the
    dimer. The solid lines correspond to $Z_{\rm el+pn}T$,
      including both the electronic and phononic contributions to the heat
      conductance, $\kappa=\kappa_{\rm el}+\kappa_{\rm pn}$, while the dashed
      lines correspond to $Z_{\rm el}T$, including only the electronic one,
      $\kappa=\kappa_{\rm el}$.}
  \label{fig-ZT-TopTop}
\end{figure}

Thus, as in the case of the blunt-sharp geometries above, the room temperature
thermal conductance is dominated by the electronic contribution in the monomer
case, while phonon transport dictates the total value of the thermal
conductance in the dimer case. This is reflected in the $Z_{\rm el}T$ and
$Z_{\rm el+pn}T$ values of these junctions, see Fig.~\ref{fig-ZT-TopTop}. For
the monomer $Z_{\rm el}T$ and $Z_{\rm el+pn}T$ are very similar in the whole
temperature range, while $Z_{\rm el+pn}T$ is 1-3 orders of magnitude smaller
than $Z_{\rm el}T$. As it is clear from the data listed in
Table~\ref{table1}, at room temperature we find indeed $Z_{\rm el+pn}T \approx
Z_{\rm el}T$ for the monomer, while for the dimer $\kappa_{\rm pn}$ reduces
$Z_{\rm el+pn}T$ as compared to $Z_{\rm el}T$ by more than one order of
magnitude. This illustrates once more the key role played by phonon transport
in these dimer junctions.

\begin{figure}[t]
\begin{center} \includegraphics[width=\columnwidth,clip]{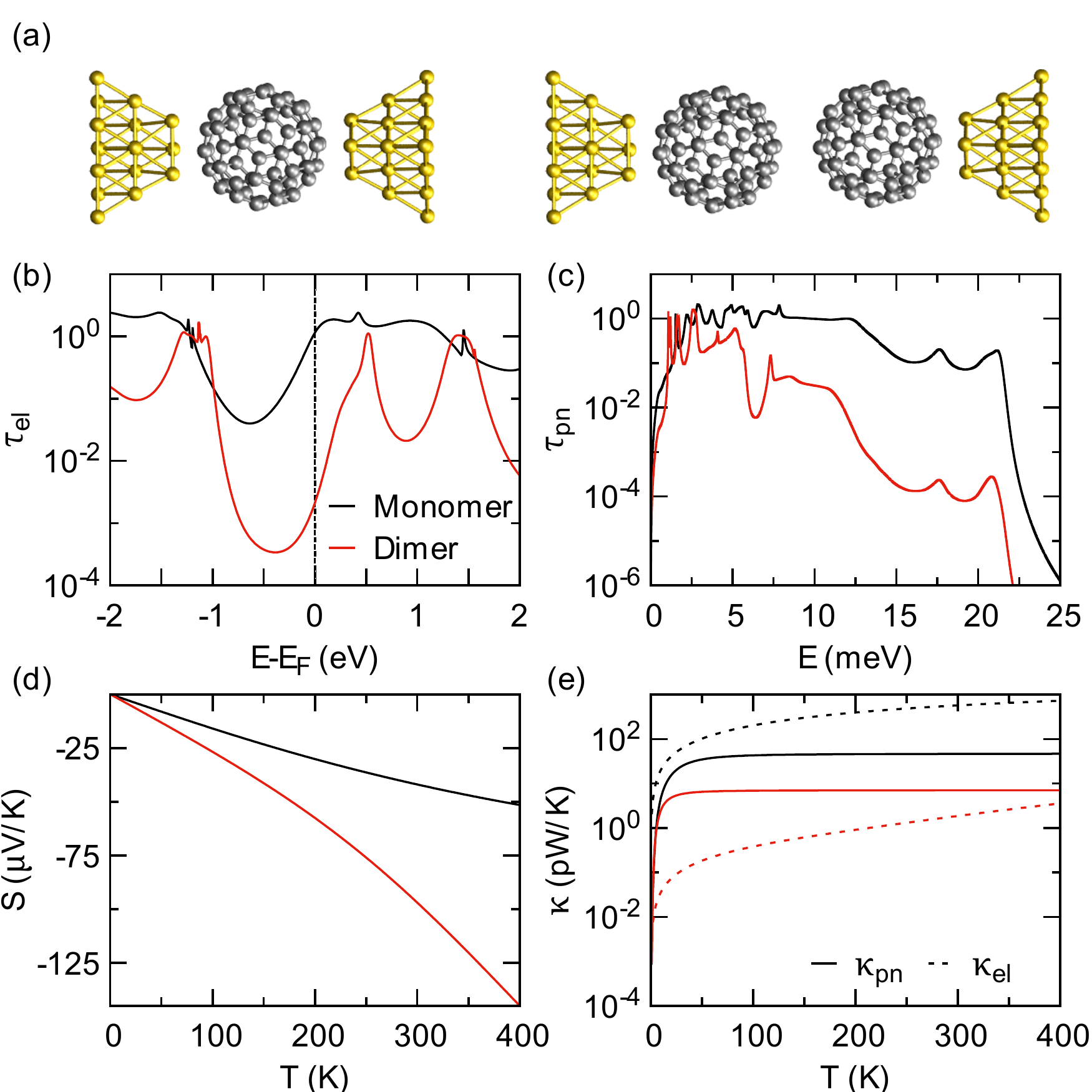} \end{center}
\caption{(Color online) (a) Equilibrium geometries of a C$_{60}$ monomer and a
  C$_{60}$ dimer, bonded to both Au electrodes through blunt
  tips. These geometries correspond to the minimum of the
    total energy with respect to the distance $d$ between the electrodes,
    i.e.\ $d=d_0$. (b) Electronic transmission as a function of energy,
  measured with respect to the Fermi level, for the two geometries shown in
  panel (a). (c) Phonon transmissions as a function of energy. (d) Thermopower
  of the two junctions as a function of temperature. (e) Electronic and
  phononic contributions to the heat conductance for monomer and dimer
  junctions as a function of temperature. Data referring to the monomer is
  shown in black, those of the dimer in red.}
\label{fig-HolHol}
\end{figure}

\begin{figure}[bt]
  \begin{center} \includegraphics[width=0.8\columnwidth,clip]{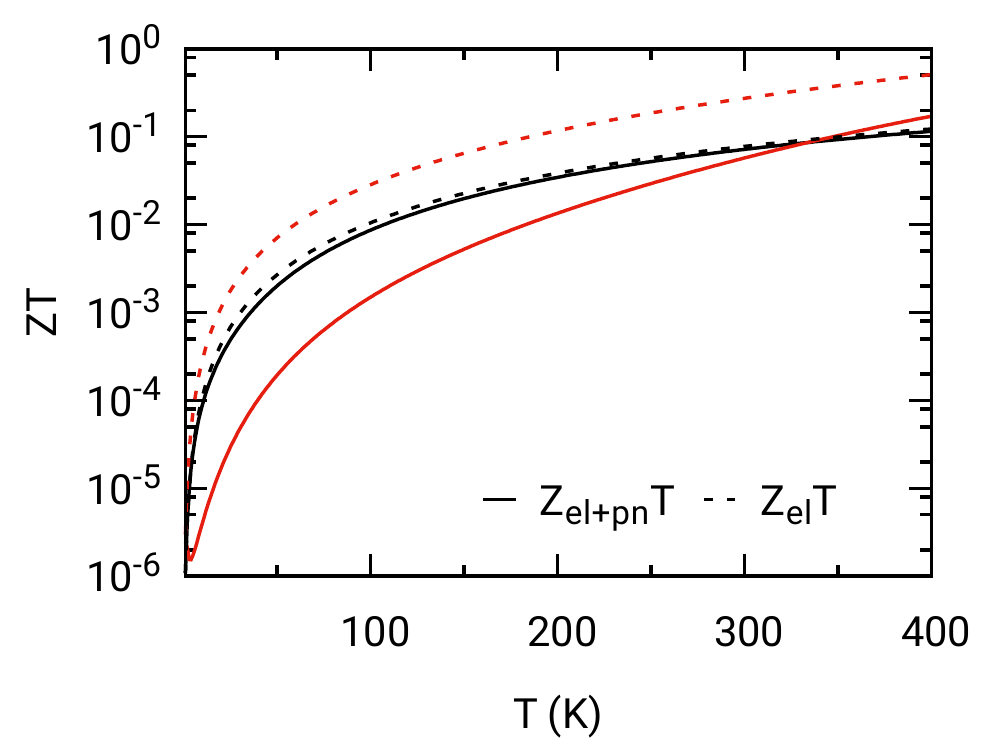} \end{center}
  \caption{(Color online) Figure of merit $ZT$ as a function of temperature
    for the two geometries shown in Fig.~\ref{fig-HolHol}(a). Black curves are
    for the monomer junction, while red ones are for the
    dimer. The solid lines correspond to $Z_{\rm el+pn}T$,
      including both the electronic and phononic contributions to the heat
      conductance, $\kappa=\kappa_{\rm el}+\kappa_{\rm pn}$, while the dashed
      lines correspond to $Z_{\rm el}T$, including only the electronic one,
      $\kappa=\kappa_{\rm el}$.}
  \label{fig-ZT-HolHol}
\end{figure}

In Figs.~\ref{fig-HolHol} and \ref{fig-ZT-HolHol} we show results for yet
another binding geometry, where in this case the molecules are bonded to both
electrodes through blunt tips, see Fig.~\ref{fig-HolHol}(a). The room
temperature values of the different thermoelectric transport properties are
summarized in Table \ref{table1}. This binding geometry allows us to test our
conclusions in a situation, where the metal-molecule coupling takes place
through several Au atoms on both sides. As one can see in these two figures,
the main conclusions of our discussions above are confirmed
again. In particular, $Z_{\rm el+pn}T$ is very similar to
  $Z_{\rm el}T$ for the monomer, since $\kappa_{\rm pn}$ is negligible as
  compared to $\kappa_{\rm el}$, while there is a strong reduction of $Z_{\rm
    el+pn}T$ as compared to $Z_{\rm el}T$ for the dimer since $\kappa_{\rm
    pn}$ dominates over $\kappa_{\rm el}$.


A major difference of the blunt-blunt geometry with respect to the previous
two types of geometries is the fact that the stronger electronic
metal-molecule coupling leads to a rather high electronic transmission at the
Fermi energy, which is reflected both in the electrical conductance and in the
electronic contribution to the thermal conductance.  As it is clear from
Table~\ref{table1}, both $G$ and $\kappa_{\rm el}$ increase monotonically for
monomer and dimer junctions in the order of sharp-sharp, blunt-sharp, and
blunt-blunt geometries. Let us emphasize that the metal-molecule binding
geometry also limits the efficiency of phonon heat transfer in the monomer
junctions, with the most efficient coupling for the blunt tips. Indeed we
obtain a clear ordering of $\kappa_{\rm pn}$, which increases from sharp-sharp
to sharp-blunt and blunt-blunt junctions. However, the phonon thermal
conductance of the dimer junctions is nearly insensitive to the metal-molecule
coupling and limited by the weak molecule-molecule coupling.

Let us conclude this section by noting that we have also simulated the
stretching of the sharp-sharp and blunt-blunt geometries (not shown here) and
found that the main conclusions also apply there. In particular, we find in
all of the cases that the phonon transport is very detrimental for the $ZT$
values of the dimer junctions, while it plays a marginal role in the case of
the monomer junctions.

\subsection{Photon transport and thermoelectric figure of merit} \label{subsec-Results-pt}

The question that we want to address in this section is whether photon
transport via thermal radiation can have an impact on the figure of merit of
molecular junctions. Due to NFRHT the thermal conductance $\kappa_{\rm pt}$
can indeed largely exceed limits set by the by the Stefan-Boltzmann law for
black bodies \cite{Song2015a,Polder1971}. We therefore put $\kappa_{\rm pt}$
in relation to $\kappa_{\rm el}$ and $\kappa_{\rm pn}$, which were discussed
before.

\begin{figure}[t]
\begin{center} \includegraphics[width=1.0\columnwidth,clip]{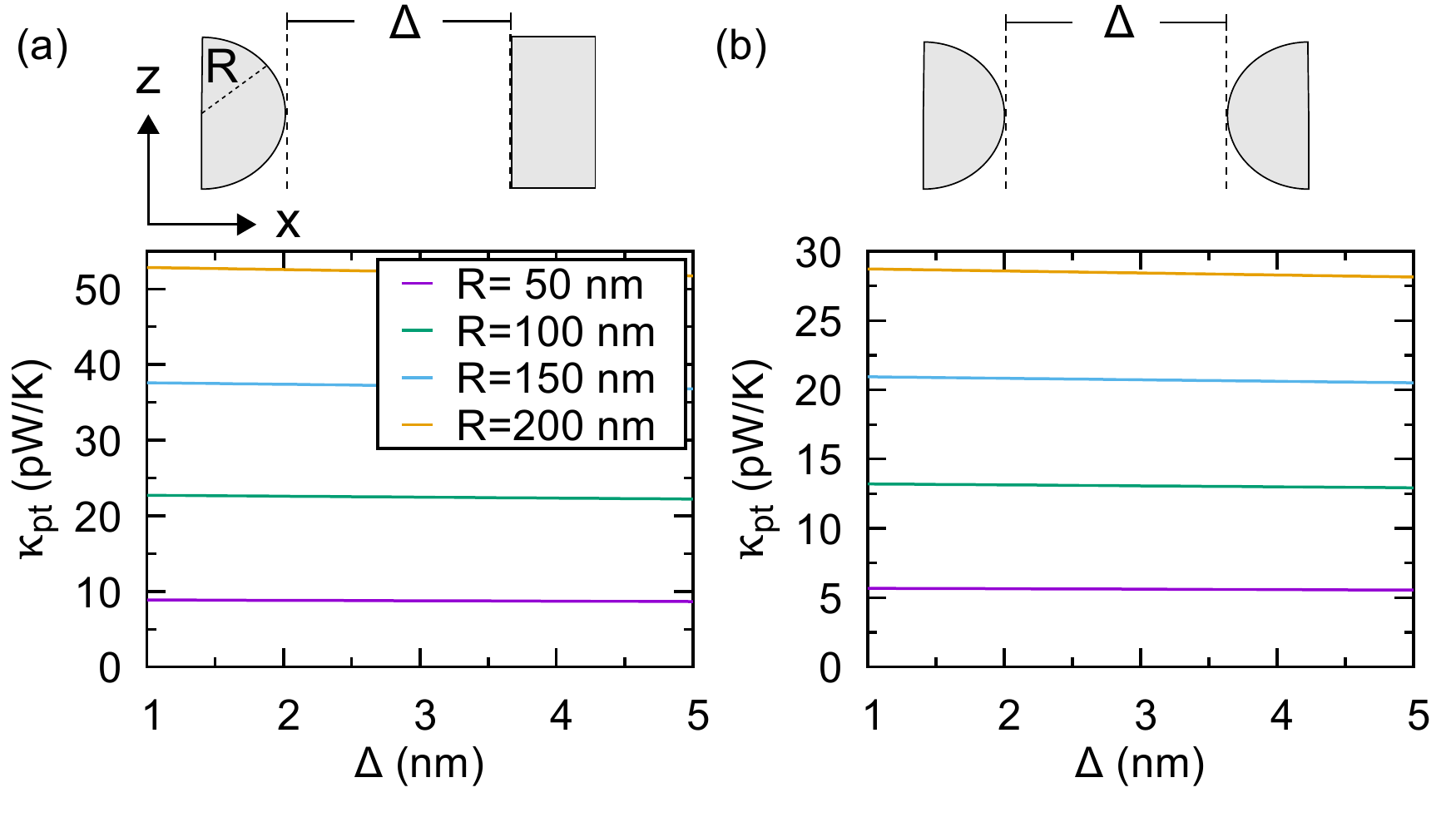} \end{center}
\caption{(Color online) Room temperature radiative heat conductance as a
  function of the gap size for (a) a tip-surface geometry and (b) a tip-tip
  geometry. The different curves corresponds to different values of the radius
  of the spheres used to model the tips in both types of geometries.}
\label{fig-RHT}
\end{figure}

As described in subsection \ref{subsec-theory-pt}, we use the formalism of
fluctuational electrodynamics \cite{Rytov1989} to treat
NFRHT. The results for the room temperature radiative heat
  conductance $\kappa_{\rm pt}$, obtained with this procedure, are shown in
  Fig.~\ref{fig-RHT} as a function of the gap size (or distance between the
  electrodes) $\Delta$ for the tip-surface and tip-tip geometries. We show
these results in a relatively large gap-size range from 1 to 5~nm to provide
an idea of the expectations for the photonic contribution to the thermal
conductance in a wide range of molecular junctions. Notice that the radiative
heat conductance changes quite slowly with the gap size.

  For our purposes, we can assume that $\kappa_{\rm pt}$
   basically remains constant in the range of studied electrode-to-electrode
   distances $\Delta$. Defined as shown in Figs.~\ref{fig-Taud} and
   \ref{fig-RHT} and listed in Table~\ref{table1}, gap sizes $\Delta$ between
   the electrodes in our molecular junction geometries vary between 1 and
   3~nm. Maximal elongations $d-d_0$, considered in Fig.~\ref{fig-Resultd},
   remain below 1~nm. The range of $\Delta$, studied in Fig.~\ref{fig-RHT}, is
   thus compatible with the atomistic molecular junction models.

To relate $\kappa_{\rm pt}$ to $\kappa_{\rm el}$ and
  $\kappa_{\rm pn}$, we consider again Table~\ref{table1}. The comparison of
  the listed thermal conductances with Fig.~\ref{fig-RHT} shows that,
  depending on the radius of the tip used to model the electrodes, the
  photonic contribution to the thermal conductance can be comparable or larger
  than the phononic one in the contact regime. In some cases $\kappa_{\rm pt}$
  can even exceed the electronic contribution $\kappa_{\rm el}$. This happens
  for both types of nanogap configurations, i.e. tip-surface and tip-tip
  geometries. It is obvious that $\kappa_{\rm pt}$ is particularly important
  in the tunneling regime. There it dominates the thermal transport, if
  junctions are stretched just by a few \AA, see Fig.~\ref{fig-Resultd}.

\begin{figure}[t]
\begin{center} \includegraphics[width=1.0\columnwidth,clip]{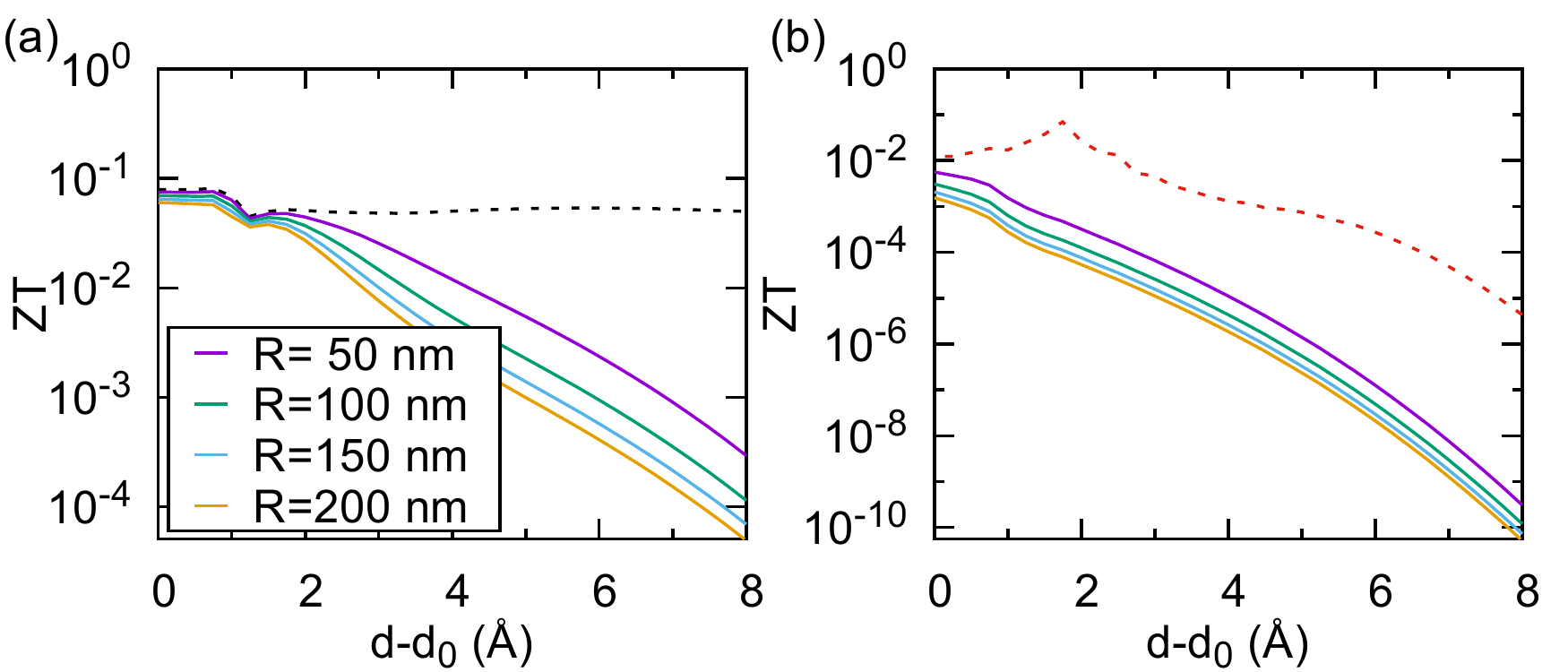} \end{center}
\caption{(Color online) Comparison of the room temperature
    figure of merit $Z_{\rm el+pn+pt}T$ (solid colored lines), including
    electronic, phononic and photonic contributions to the thermal
    conductance, with $Z_{\rm el+pn}T$ (dashed lines), including only
    electronic and phononic parts, as a function of elongation for (a) the
    hollow-top monomer junction and (b) the hollow-top dimer junction. The
  dashed curves for $Z_{\rm el+pn}T$ are identical to those shown in
  Fig.~\ref{fig-Resultd}(d) and are reproduced to provide a reference. To
  determine the photonic heat conductance, the tip-surface geometry was
  assumed with tips of different radii $R$, as indicated by the legend.}
\label{fig-ZT-elpnpt}
\end{figure}

To illustrate the impact of $\kappa_{\rm pt}$ on the figure of merit, we
display in Fig.~\ref{fig-ZT-elpnpt} the $ZT$ values as a function of the
displacement of the electrodes for the blunt-sharp junctions considered in
Fig.~\ref{fig-Resultd}. In particular, we show how the $ZT$ values are
modified by the photonic contribution for the tip-surface configuration and
for various values of the tip radius. Notice that for
  elongations of the electrodes off the equilibrium by up to 2 \AA, $Z_{\rm
    el+pn+pt}T$ with $\kappa=\kappa_{\rm el}+\kappa_{\rm pn}+\kappa_{\rm pt}$
  for the monomer junction is similar to $Z_{\rm el+pn}T$ that considers only
  electronic and phononic contributions. But for larger separations, $Z_{\rm
  el+pn+pt}T$ breaks down dramatically.  For the dimer
  junction, a suppression of $Z_{\rm el+pn+pt}T$ as compared to $Z_{\rm
    el+pn}T$ is obvious throughout the full range of elongations $d-d_0$
  considered. Generally speaking, the detrimental influence of $\kappa_{\rm
  pt}$ on $ZT$ is strongest in the tunneling regime for large electrode
separations.

\section{Conclusions} \label{sec-Conclusions}

In summary, we have presented a systematic \emph{ab initio} study of the role
of the phonon transport in the thermal conductance and thermoelectric figure
of merit of C$_{60}$-based molecular junctions. In particular, we have
analyzed both monomer and dimer junctions, where the fullerenes are attached
to gold electrodes. Taking only electrons and phonons into account, we have
found that the thermal transport of monomer junctions is dominated by the
electronic contribution irrespective of the binding geometry and elongation
stage, while in the case of dimer junctions phonons dominate over
electrons. This latter fact has important consequences for the thermoelectric
figure of merit of dimer junctions, which is significantly reduced by the
phononic thermal conductance. Our findings suggest that the proposal that
stacks of C$_{60}$ molecules could constitute a strategy to realize very
efficient thermoelectric molecular devices is not justified.

On the other hand, we have also analyzed the importance of near-field thermal
radiation for the figure of merit of molecular junctions. We have shown that,
depending on the geometry of the molecular junction, the photonic contribution
to the thermal conductance can be very significant and, in turn, detrimental
for the performance of molecular junctions as thermoelectric devices. The
photonic contribution is increasingly relevant in the tunneling regime, when
covalent bonds to one of the electrodes are broken.

Overall our work sheds new light on the fundamental role of phonon and photon
transport in the thermal conduction properties of molecular junctions. It
shows the critical importance of understanding all factors determining the
heat transport in order to assess the performance of these nanojunctions as
potential energy conversion devices. We expect that our predictions can soon
be verified experimentally by using techniques similar to those developed in
Ref.~\onlinecite{Cui2017a}.

\section{Acknowledgments}

J.C.K.\ and F.P.\ gratefully acknowledge funding from the Carl Zeiss
foundation and the Junior Professorship Program of the Ministry of Science,
Research, and the Arts of the state of Baden W\"urttemberg. J.C.C.\ was
supported through the Spanish Ministry of Economy and Competitiveness
(Contract No.\ FIS2014-53488-P) and thanks the German Research Foundation
(DFG) and the Collaborative Research Center (SFB) 767 for sponsoring his stay
at the University of Konstanz as Mercator Fellow. An important part of the
numerical modeling was carried out on the computational resources of the bwHPC
program, namely the bwUniCluster and the JUSTUS HPC facility.


\end{document}